\documentclass[preprint,showpacs,preprintnumbers]{revtex4}
\usepackage{amsmath}
\usepackage{graphicx}
\usepackage{dcolumn}
\usepackage{bm}
\usepackage{epsfig}
\usepackage[T1]{fontenc}
\usepackage{ae,aecompl}

\begin{document}

\title{Phase separation in thermal systems: LB study and morphological
characterization}
\author{Yanbiao Gan$^{1,2,3}$, Aiguo Xu$^2$\footnote{
Corresponding author. E-mail: Xu\_Aiguo@iapcm.ac.cn}, Guangcai
Zhang$^2$, Yingjun Li$^1$\footnote{Corresponding author. E-mail:
lyj@aphy.iphy.ac.cn}} \affiliation{1, State Key Laboratory for
GeoMechanics and Deep Underground Engineering, SMCE, China
University of Mining and Technology (Beijing), Beijing 100083,
P.R.China\\
2, National Key Laboratory of Computational Physics, \\
Institute of Applied Physics and Computational Mathematics, P. O. Box
8009-26, Beijing 100088, P.R.China \\
3, North China Institute of Aerospace Engineering, Langfang 065000,
P.R.China }
\date{\today }

\begin{abstract}
We investigate thermal and isothermal symmetric liquid-vapor
separations via an FFT-thermal lattice Boltzmann (FFT-TLB) model.
Structure factor, domain size, and Minkowski functionals are
employed to characterize the density and velocity fields, as well as
to understand the configurations and the kinetic processes. Compared
with the isothermal phase separation, the freedom in temperature
prolongs the spinodal decomposition (SD) stage and induces different
rheological and morphological behaviors in the thermal system. After
the transient procedure, both the thermal and isothermal separations
show power-law scalings in domain growth; while the exponent for
thermal system is lower than that for isothermal system. With
respect to the density field, the isothermal system presents more
likely bicontinuous configurations with narrower interfaces, while
the thermal system presents more likely configurations with
scattered bubbles. Heat creation, conduction, and lower interfacial
stresses are the main reasons for the differences in thermal system.
Different from the isothermal case, the release of latent heat
causes the changing of local temperature which results in new local
mechanical balance. When the Prandtl number becomes smaller, the
system approaches thermodynamical equilibrium much more quickly. The
increasing of mean temperature makes the interfacial stress lower in
the following way:
$\sigma=\sigma_{0}[(T_{c}-T)/(T_{c}-T_{0})]^{3/2}$, where $T_{c}$ is
the critical temperature and $\sigma_{0}$ is the interfacial stress
at a reference temperature $T_{0}$, which is the main reason for the
prolonged SD stage and the lower growth exponent in thermal case.
Besides thermodynamics, we probe how the local viscosities influence
the morphology of the phase separating system. We find that, for
both the isothermal and thermal cases, the growth exponents and
local flow velocities are inversely proportional to the
corresponding viscosities. Compared with isothermal case, the local
flow velocity depends not only on viscosity but also on temperature.
\end{abstract}

 \maketitle
\preprint{APS/123-QED}
\date{\today}

\section{Introduction}

Multiphase flows and heat transfers are ubiquitous in natural,
industrial processes, as well as daily life, e.g., oil-water
systems, bubble flows, petroleum processing, paper-pulping, and
power plants, etc \cite{Fun-Multiphase}. Therefore, establishing
accurate prediction models to investigate the underlying
 physical essence of these phenomena, is of great academic significance
and industrial practical value. However, due to the complex nature
and inherent nonlinearities of multiphase flows, theoretical
solutions are usually limited to a small class of problems in
one-dimension and with numerous simplifying assumptions and
generalizations \cite{Exact solution}. On the other hand,
experimental approaches for multiphase flows are generally expensive
and some problems are still being unsolved in accurate measurement
technology (e.g., interfacial area measurement) for this process
\cite{Exp-Area}. Consequently, it is reasonable to consider
numerical simulation, to some extent, as a primarily useful tool in
studying the underlying physics of multiphase flows and providing
some insights into understanding the kinetic process, that are
difficult to obtain from theoretical analysis or experiments.

Molecular dynamics (MD) is a nice microscopic approach, but it is
too computationally expensive to access dynamic behaviors with
spatiotemporal scales comparable with experiments
\cite{Succi-PRL-2006}. Moreover, many macroscopic behaviors are, in
fact, not sensitive to degrees of freedom at the molecular level.
Traditional fluid dynamics does not work well for systems where
non-equilibrium effects are pronounced, for example, multiphase
system. In addition, from the computational expenses point of view,
the direct simulation of fluid behaviors in such a system is also a
challenging work, since it is not easy to track the deformable
macroscopic interfaces and the incorporate complex microscopic
interactions \cite{GZL-03PRE}.

Between these two approaches, as a mesoscopic approach, the lattice
Boltzmann (LB) method, has enjoyed substantial development and has
become a very promising and versatile tool for simulating complex
phenomena in various fields during the past two decades
\cite{Succi-Book}, ranging from magnetohydrodynamics
\cite{Succi-PRA-1991,sofonea-EPJB-1997}, to compressible flows
\cite{Watari-Tsutahara,Xu-compressible-1,Xu-compressible-2}, wave
propagations \cite{wave}, hydrodynamic instabilities
\cite{JCP-1999,KHI}, etc. Apart from fields listed above, the
versatile method is particularly promising in the area of multiphase
systems \cite{RK-model,shanchen,Yeomans,contact-line-motion-2004,
contact angle shanchen, SC-model--wetting-PRE-2006,
SC-model-nonwetting-PRE-2007, 3D-DROPLET BREAK, break-pre-2002,
POF-PRE-2005-Abraham,reactive-Yeomans-2006,XGL1,Sofonea-multiphase-pre-2004,
He-JSP-2002,Shan-PRE-2006,Yuan-POF-2006,Lee-PRE-2006,Seta-JFST-2007,Sbragaglia-JFM-2009}%
. This is mainly owing to its intrinsic kinetic nature, which makes
the inter-particle interactions (IPI) be incorporated easily and
flexibly and, in fact, the IPI is the underlying microscopic
physical reason for phase separation and interfacial tension in
multiphase systems. So far, many LB models for multiphase flows have
been proposed, among which the three well-known ones are the
Chromodynamic model by Gunstensen \emph{et al}. \cite{RK-model}, the
pseudo-potential model by Shan and Chen
\cite{shanchen,Shan-PRE-2006}, and the free energy model by Swift
\emph{et al}. \cite{Yeomans}.

The aforementioned models have been successfully applied to study a
wide variety of multiphase flow problems in science and engineering,
such as contact line motion \cite{contact-line-motion-2004, contact
angle shanchen}, wetting
\cite{SC-model--wetting-PRE-2006,SC-model-nonwetting-PRE-2007}, drop
breakup \cite{3D-DROPLET BREAK,break-pre-2002}, drop collision \cite%
{POF-PRE-2005-Abraham}, chemically reactive fluid \cite%
{reactive-Yeomans-2006}, phase separation and phase ordering
\cite{RK-model,shanchen,Yeomans,XGL1,Sofonea-multiphase-pre-2004,Seta-JFST-2007},
etc. Despite this, to date, most studies focus on the isothermal
systems, because, in these models, only mass and momentum
conservations are kept, hydrodynamic behaviors due to temperature
field are not taken into account. However, thermal effects are
significant, even dominant, in many cases. Examples are referred to
phase separations in boiling process, distillation and condensation
process, and thermal nuclear reactor, etc. In these systems, the
evolutions of the temperature field and flow field is spontaneously
coupled with each other \cite{Onuki-book,onuki}. Therefore, it is a
fundamental and essential work to develop thermal LB (TLB) models
for multiphase system. But due to the complexity of this problem,
the progress has been rather slow.

The most obvious obstacle lies in the fact that, when the
interparticle forces are incorporated, how to ensure the total
energy conservation becomes challenging in the discrete model. To
overcome this difficulty, extensive efforts have been made over the
past years. But until very recently, only a few TLB models for
multiphase flows have been proposed, and can be roughly divided into
two approaches. The first is the passive scalar approach \cite
{ZRY-2003PRE,YuanPeng-DrThesis}. In this approach, evolutions of the
density field and the momentum field are solved by an isothermal LB
model, while the temperature evolution is determined by an
additional passive-scalar equation. The coupling of these two parts
is through a suitably defined body force in the isotherm LB
equation. This approach is conceptually rather simple and as stable
as the isothermal LB models, because the energy conservation is not
explicitly implemented. Meanwhile, it can produce a non-ideal gas
equation of state (EOS) and capture the temperature field. However,
it should be pointed out that, in the passive scalar approach, the
viscous dissipation and compression work done by the pressure are
neglected \cite{YuanPeng-DrThesis}.

The second is the multispeed approach, which implements energy
conservation by using larger and more isotropic sets of velocities
and by including higher order velocity terms in the equilibrium
distribution. Examples for ideal gas include the works of Alexander
\emph{et al.} \cite{Alexander-PRE-1993}, Watari \emph{et al.}
\cite{Watari-Tsutahara}, Xu \emph{et al.}
\cite{Xu-compressible-1,KHI}, and so on. However, applications of
this approach for thermal flows with high Mach number or flows with
high Knudsen number still have some challenges. The challenges arise
from the insufficient truncation in the equilibrium distribution
function and the insufficient isotropy in the
discrete-velocity-model (DVM). In an alternative way, using the
Hermite expansion approach, Shan \emph{et al.} \cite{Shan-JFM-2006}
presented a systematic theoretical framework for constructing TLB
models that approximate the continuum Boltzmann equation with higher
accuracy. With the Hermite expansion approach, hydrodynamic moments
at various levels can be determined in a straightforward way at a
given order of truncations of the Hermite polynomials. Almost
simultaneously, similar results were obtained by Philippi \emph{et
al.} \cite{Philippi-PRE-2006} using a different procedure. Although
the above-mentioned TLB models only work for ideal gas systems, they
can be extended to multiphase flows by the extra force method. The
one developed by Gonnella, Lamura, and Sofonea (GLS)
\cite{GLS-model-PRE2007} is typical. In this model, an extra term
$I_{ki}$, accounting for inter-particle forces, is added into the LB
equation to describe the van der Waals (VDW) fluids. From the point
view of the IPI, it can be considered as a bottom-up approach,
similar to the Shan-Chen model. To describe system with interfaces,
gradient contributions to free energy due to the inhomogeneity of
fluid density are also included. Compared with the passive scalar
approach, all observable fields, e.g., density, velocity,
temperature, and pressure are directly derived from the same
distribution function, as in the standard kinetic theory.

In a recent work \cite{PLA}, we further develop GLS model so that
the total energy conservation can be better held and the spurious
velocities can be damped to negligible scale in the numerical
simulations. In the improved model, spatial derivatives in the
convection term and the force term are calculated via the fast
Fourier transformation (FFT) and its inverse (IFFT). For convenience
of description, we refer to this model as FFT-TLB model. Via the
FFT-TLB model, we study the effects of temperature and viscosity on
liquid-vapor phase separation in two-dimensional case. It is known
that, spatial domains of homogeneous phases evolving during spinodal
decomposition (SD) show a large variety of complex spatial patterns
and the system is globally in a nonequilibrium state. How to
effectively describe and pick up information from such a complex
system is still an open problem. In the present work, besides the
rheological behavior, we use the Minkowski functionals
\cite{Minkowski functionals} to characterize the isothermal and
thermal phase separations and conduct a comparison study on the
similarities and differences between these two cases.

The following part of the paper is planned as follows. The Minkowski
functionals and the FFT-TLB model are briefly reviewed in Secs. II and III,
respectively. Simulation results and corresponding physical interpretations
are given in Sec. IV. Sec. V presents conclusions and discussions.

\section{Morphological characterization}

In this section, we briefly review the set of statistics known as
Minkowski functionals \cite{Minkowski functionals}, which will be
used to characterize the physical fields in Sec. IV. Such a
description has been well known in digital picture analysis
\cite{digital picture analysis} and
successfully adapted to characterize the reaction-diffusion systems \cite%
{reaction}, shocked porous materials \cite{xu-porous material}, and
patterns in phase separation of complex fluids
\cite{JCP,sofonea-PRE-1997}, etc.

According to a general theorem of integral geometry, all properties of a $d$%
-dimensional convex set, which satisfy motion invariance and additivity, are
contained in $d+1$ numerical values \cite{D+1}. For a pixelized map $\psi(%
\mathbf{x})$, we consider the excursion sets of the map, defined as
the set of all map pixels with value of $\psi$ greater than some
threshold $\psi_{th} $, where $\mathbf{x}$ is the position, $\psi$
can be a state variable like density $\rho$, temperature $T$, or
pressure $P$; $\psi$ can also be the velocity $\mathbf{u}$ or its
components, or some specific stress, etc. Then the $d+1$ functionals
of these excursion sets completely describe the morphological
properties of the underlying map $\psi(\mathbf{x})$. In the case of
two- or three-dimensions, the Minkowski functionals have intuitive
geometric interpretations. For a two-dimensional density map $\rho(\mathbf{x}%
)$, the three Minkowski functionals correspond geometrically to the
fractional area $A$ of the high density domains, the boundary length $L$
between the the high and low density domains, and the Euler characteristic $%
\chi$.

In this work, we probe the effects of temperature and velocity on
phase separation by checking the density map $\rho(\mathbf{x},t)$
and velocity map $\mathbf{u}(\mathbf{x},t)$, where time $t$ is
explicitly denoted. When the density $\rho(\mathbf{x},t)$ is beyond
the threshold value $\rho_{th}$, the grid node at position
$\mathbf{x}$ is regarded as a white vertex, otherwise it is regarded
as a black one. For the square lattice, a pixel possesses four
vertices. A region with connected white (black) pixels is defined as
a white (black) domain. Two neighboring white and black domains
present an interface or boundary. When the threshold contour level
$\rho_{th}$ increase from the lowest density $\rho_{\min}$ to the
highest one $\rho_{\max}$, the white area fraction $A=N_{A}^{w}/N$
will decrease from $1$ to $0$, and the qualitative features of the
patterns will vary drastically, where $N_{A}^{w}$ is the number of
pixels with a density larger than $\rho_{th}$, $N=N_{x}\times N_{y}$
is the total number of pixels, $N_{x}$ and $N_{y}$ are the lattice
numbers along the $x$ and $y$ directions; the boundary length
$L=N_{L}/N$ is defined as the ratio between the pixels separating
the black and white domains, and the total number of pixels. With
the increasing of $\rho_{th}$, boundary length $L$ first increases
from $0$ at $\rho_{th}=\rho_{\min}$, then arrives at a maximum value
$L_{\max}$ and, finally decreases to $0$ again at
$\rho_{th}=\rho_{\max}$; the third morphological quantity is the
Euler characteristic $\chi$, defined as the difference of the number
of connected white domains $N_{\chi}^{w}$ and black
domains $N_{\chi}^{b}$ normalized by $N$, $\chi=(N_{\chi}^{w}-N_{%
\chi}^{b})/N $. In contrast to the white area $A$ and boundary length $L$,
the Euler characteristic $\chi$ describes the connectivity of the domains in
a purely topological way. It is negative (positive) if many disconnected
black (white) regions dominate the image. A vanishing Euler characteristic
indicates a highly connected structure with equal numbers of black and white
domains. Despite having global meaning, the Euler characteristic $\chi$ can
be calculated in a local way using the additivity relation \cite%
{reaction,sofonea-PRE-1997}. Since the measures are normalized by $N$, they
can be used to compare systems with different sizes.

\section{FFT-TLB multiphase model}

In this section, we present the FFT-TLB model for simulating thermal
liquid-vapor system. The model is a further development of the one
proposed by GLS \cite{GLS-model-PRE2007}. GLS introduced an
appropriate inter-particle force term to describe the VDW fluids.
Our contribution is to propose an appropriate FFT scheme, which is
used to calculate the convection term and the force term. With this
new model, the non-conservation problem of total energy due to
spatiotemporal discretizations is much better solved and spurious
currents in equilibrium interfaces are significantly reduced in the
numerical simulations.

\subsection{TLB multiphase model by GLS}

The GLS model includes the following two parts: (i) TLB model by
Watari-Tsutahara (WT) \cite{Watari-Tsutahara}; (ii) an appropriate
inter-particle force, $I_{ki}$. The original WT model works only for
ideal gas. It uses the following DVM:
\begin{equation}
\mathbf{v}_{0}=0\text{, }\mathbf{v}_{ki}=v_{k}[\cos (\frac{i-1}{4}\pi )\text{%
, }\sin (\frac{i-1}{4}\pi )]\text{, }k=1,2,3,4\text{; }i=1,2...8\text{,}
\end{equation}%
where subscript $k$ indicates the $k$-th group of particle
velocities whose speed is $v_{k}$ and $i$ indicates the direction of
particle's speed. Different from the standard LB model, WT model
uses a second upwind finite-difference (FD) scheme to calculate the
convection term in the LB equation. The FD LB model breaks the
combination of discretizations of space and time, which makes the
particle speeds more flexible. The values of the speeds
$\mathbf{v}_{k}$ may be determined in such a way that the
temperature gets a large interval around the critical temperature
$T_{c}$, under which the simulation is stable. This is of great
importance for phase separation studies where long lasting
simulations are needed to determine the growth behavior \cite{XGL1}.

Compared to WT model, the main contribution of GLS model is the
introduction of the extra term $I_{ki}$, which accounts for
inter-particle forces
\begin{equation}
\frac{\partial f_{ki}}{\partial t}+\mathbf{v}_{ki}\cdot \frac{\partial f_{ki}%
}{\partial \mathbf{r}}=-\frac{1}{\tau }\left[ f_{ki}-f_{ki}^{eq}\right]
+I_{ki}\text{,}  \label{Iki_bgk}
\end{equation}
where $f_{ki}^{eq}$ is the local equilibrium distribution function; $\mathbf{%
r}$ is the spatial coordinate; $\tau $ is the relaxation time
related to the kinematic viscosity. The distribution function
$f_{ki}^{eq}$ is related to the local density $\rho $, fluid
velocity $\mathbf{u}$, and temperature $T$ through the following
moments:
\begin{equation}
\rho =\sum_{ki}f_{ki}^{eq}\text{,}  \label{n_eq}
\end{equation}%
\begin{equation}
\rho \mathbf{u}=\sum_{ki}\mathbf{v}_{ki}f_{ki}^{eq}\text{,}  \label{nu_eq}
\end{equation}%
\begin{equation}
\rho T=\sum_{ki}\frac{1}{2}(\mathbf{v}_{ki}-\mathbf{u})^{2}f_{ki}^{eq}\text{.%
}  \label{p_eq}
\end{equation}
$I_{ki}$ in Eq. (\ref{Iki_bgk}) takes the following form:
\begin{equation}
I_{ki}=-[A+B_{\alpha }(v_{ki\alpha }-u_{\alpha })+(C+C_{q})(v_{ki\alpha
}-u_{\alpha })^{2}]f_{ki}^{eq},  \label{iki}
\end{equation}%
with
\begin{equation}
A=-2(C+C_{q})T,  \label{AAA}
\end{equation}%
\begin{equation}
B_{\alpha }=\frac{1}{\rho T}[\partial_{\alpha }(P^{w}-\rho
T)+\partial_{\beta }\Lambda_{\alpha\beta }-\partial _{\alpha }(\zeta
\partial_{\gamma}u_{\gamma })],  \label{BBB}
\end{equation}%
\begin{eqnarray}
C&=&\frac{1}{2\rho T^{2}}\{(P^{w}-\rho T)\partial _{\gamma }u_{\gamma
}+\Lambda _{\alpha \beta }\partial_{\alpha }u_{\beta }-(\zeta \partial
_{\gamma }u_{\gamma })\partial _{\alpha }u_{\alpha }  \notag \\
&&\text{}+\frac{9}{8}\rho ^{2}\partial _{\gamma }u_{\gamma }+K[-\frac{1}{2}%
(\partial _{\gamma }\rho )(\partial _{\gamma }\rho )(\partial _{\alpha
}u_{\alpha })  \notag \\
&&\text{}-\rho (\partial_{\gamma }\rho )(\partial _{\gamma }\partial
_{\alpha }u_{\alpha })-(\partial _{\gamma }\rho )(\partial _{\gamma
}u_{\alpha })(\partial_{\alpha }\rho )]\},  \label{ss}
\end{eqnarray}%
\begin{equation}
C_{q}=\frac{1}{2\rho T^{2}}\partial _{\alpha }[2q\rho T(\partial _{\alpha
}T)].  \label{CCQQ}
\end{equation}%
$P^{w}=3\rho T/(3-\rho )-9\rho ^{2}/8$ is the VDW EOS. Since the pressure is
not monotonic in density, thermodynamic phase transition may occur in such a
system. By setting $\partial P^{w}/\partial \rho =0$, $\partial
^{2}P^{w}/\partial \rho ^{2}=0$, we obtain the critical point $%
T_{c}=\rho _{c}=1$. $\Lambda _{\alpha \beta }=M\partial _{\alpha }\rho
\partial _{\beta }\rho -[\rho T\partial _{\gamma }\rho \partial _{\gamma
}(M/T)]\delta _{\alpha \beta }-M(\rho \nabla ^{2}\rho +\left\vert
\nabla \rho \right\vert ^{2}/2)\delta _{\alpha \beta }$ is the
contribution of density gradient to pressure tensor and $M=K+HT$
allows a dependence of the surface tension on temperature, where $K$
is the surface tension coefficient and $H$ is a constant. It is
worth pointing out that, in this model, the Prandtl number $\Pr
=\eta /\kappa _{T}=\tau /2(\tau -q)$ can be changed by adjusting the
parameter $q$ in the term $C_{q}$.

It has been shown that \cite{GLS-model-PRE2007}, under the
Chapman-Enskog expansion, the above LB model recovers the following
equations for VDW fluids:
\begin{equation}
\partial _{t}\rho +\partial _{\alpha }(\rho u_{\alpha })=0,  \label{ns_eq_1}
\end{equation}
\begin{equation}
\partial _{t}(\rho u_{\alpha })+\partial _{\beta }(\rho u_{\alpha }u_{\beta
}+\Pi _{\alpha \beta }-\sigma _{\alpha \beta })=0,  \label{ns_eq_22}
\end{equation}
\begin{equation}
\partial _{t}e_{T}+\partial _{\alpha }[e_{T}u_{\alpha }+(\Pi _{\alpha \beta
}-\sigma _{\alpha \beta })u_{\beta }-\kappa _{T}\partial _{\alpha }T]=0,
\label{ns_eq_3}
\end{equation}%
where $\Pi _{\alpha \beta }=P^{w}\delta _{\alpha \beta }+\Lambda
_{\alpha \beta }$ is the non-viscous stress, and $\sigma _{\alpha
\beta }=\eta (\partial _{\alpha }u_{\beta }+\partial_{\beta
}u_{\alpha }-\partial _{\gamma }u_{\gamma }\delta _{\alpha \beta
})+\zeta \partial _{\gamma }u_{\gamma }\delta _{\alpha \beta }$ is
the dissipative tensor with the shear and bulk viscosities $\eta $
and $\zeta $. $e_{tol}=\rho T-9\rho ^{2}/8+K\left\vert \nabla \rho
\right\vert ^{2}/2+\rho u^{2}/2$ is the total energy density. It
should be mentioned that the force term also accounts for the
potential energy $-9/8\rho^2$ and interfacial energy $K\left\vert
\nabla \rho \right\vert^{2}/2$, which are sources of the kinetic
energy.

\subsection{Our contribution: spatial discretization with FFT}

In this subsection, we will review our improvements to the TLB
multiphase model: spatial derivatives in the convection term
$\mathbf{v}_{ki}\cdot \partial f_{ki}/\partial\mathbf{r}$ and in the
external force term $I_{ki}$ are calculated via the FFT scheme and
its inverse.

To illustrate the necessity, we present simulation results for a thermal
phase separation process by various numerical schemes. Here the time
derivative is calculated using the first-order forward Euler FD scheme.
Spatial derivatives in $I_{ki}$ are calculated using the second-order
central difference (2nd-CD) scheme. Spatial derivatives in convection term $%
\mathbf{v}_{ki}\cdot \partial f_{ki}/\partial \mathbf{r}$ are
calculated using the the 2nd-CD scheme, the Lax-Wendroff (LW)
scheme, the non-oscillatory and non-free-parameter dissipation (NND)
scheme \cite{zhanghanxin}, and the fifth-order weighted essentially
non-oscillatory scheme (5th-WENO) \cite{WENO}, respectively. As a
result, we find that the total energy density $e_{tol}(t)$ is not
conservative in simulations, even though it is in theoretical
analysis (see Fig. 1). The non-conservation of energy is caused by
errors of spatiotemporal discretizations.

\begin{figure}[tbp]
\center{\epsfig{file=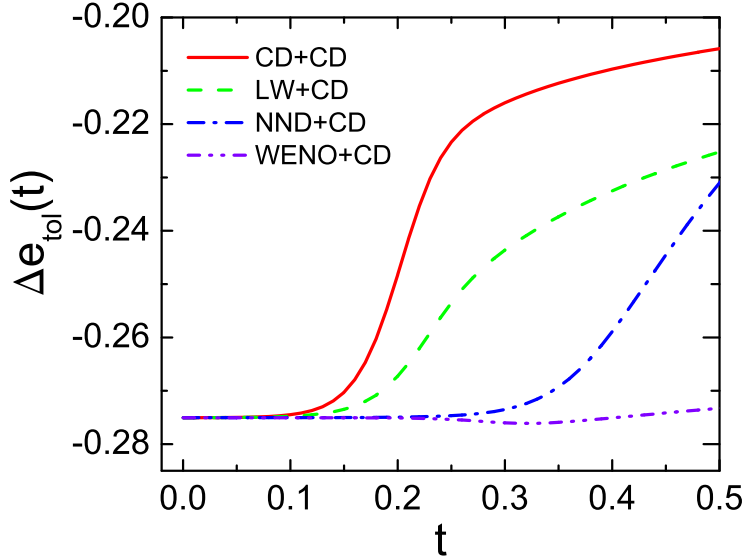,bbllx=3pt,bblly=3pt,bburx=571pt,bbury=440pt,
width=0.6\textwidth,clip=}} \caption{(Color online) Variations of
total energy density $\Delta e_{tol}(t)=e_{tol}(t)-e_{tol}(0)$ for a
phase-separating system obtained from the GLS
model with various schemes. Initial conditions are set as $\protect\rho %
=1+\Delta $, $T=0.85$, $u_{x}=u_{y}=0$, where $\Delta $ is a random
density with an amplitude of $0.01$. The remaining parameters are as
follows: $v_{1}=1.00$, $v_{2}=1.90$, $v_{3}=2.90$, $v_{4}=4.30$,
$\protect\tau =10^{-2}
$, $dx=dy=1/256$, $dt=10^{-5}$, $K=5\times 10^{-6}$, $H=0$, $\protect\zeta =0$, $%
q=-0.004$. Periodical boundary conditions (PBC) are imposed on $x$ and $y$
directions.}
\end{figure}

Aiming to solve the problem of energy non-conversation, we proposed
a new algorithm based on FFT and its inverse \cite{PLA}. This
approach is especially powerful for periodic system and also
provides spatial spectral information on field quantities. For
completeness, let us start with the definition of Fourier transform
of a discrete function $f(x_{j})$
\begin{equation}
\widetilde{f}(k)=\Delta x\sum_{j=0}^{N-1}f(x_{j})e^{-\mathbf{i}kx_{j}}\text{,%
}  \label{FT}
\end{equation}%
and its inverse
\begin{equation}
f(x_{j})=\frac{1}{L}\sum_{n=-N/2}^{N/2-1}\widetilde{f}(k)e^{\mathbf{i}kx_{j}}%
\text{,}  \label{IFT}
\end{equation}%
In Eq. (\ref{IFT}), $k\mathbf{=}2\pi n/L$ and $L=N\Delta x$ is the length of
the system divided into $N$ equal segments. A general theorem of derivative
based on FFT states that \cite{sxwlff,Spectral-Methods-Book,Plasma-Book}
\begin{equation}
\widetilde{f^{\prime }}(k)=\textbf{i} k\times
\widetilde{f}(k)\text{,} \label{FFT}
\end{equation}
where $\widetilde{f^{\prime}}(k)$ is the Fourier transform of
$f^{\prime }(x_{j})$, $k$ is the module of wave vector $\mathbf{k}$,
and $\mathbf{i}$ is an imaginary unit. The theorem provides a way to
calculate the spatial derivative $f^{\prime }(x_{j})$, composed
of the following steps: (i) transform $f(x_{j})$ in real space into $\widetilde{f}%
(k)$ in reciprocal space; (ii) multiply $\widetilde{f}(k)$ with $\mathbf{i}%
k$; (iii) take the inverse Fourier transform (IFT) of $\widetilde{%
f^{\prime }}(k)$, then the spatial derivative $f^{\prime }(x_{j})$
can be obtained. Higher-order derivative, such as the $n$th derivative $%
f^{(n)}(x_{j})$ ($n\geq 2$), can be obtained from a similar
procedure only if we multiply $\widetilde{f}(k)$ with
$(\textbf{i}k)^{n}$,
\begin{equation}
\widetilde{f^{(n)}}(k)=(\textbf{i} k)^{n}\times
\widetilde{f}(k)\text{.} \label{NFFT}
\end{equation}
High order derivatives can be calculated from this convenient way is
a main merit of FFT over FD schemes, otherwise, we should choose
more stencils (more points) to approximate high order derivatives.

The FFT approach has excellent accuracy properties, typically well
beyond that of standard discretization schemes. In principle, it
gives the exact derivative with infinite order accuracy if the
function is infinitely differentiable
\cite{Spectral-Methods-Book,Plasma-Book,Spectral-Methods-Book-2,PRL-FFT-1971}.
In our manuscript, using this virtue, the FFT scheme is designed to
approximate the true spatial derivatives, as a result, to eliminate
spurious velocities near the interface region and to guarantee
energy conservation. However, the trouble in proceeding in this
manner is that, in many cases, it is difficult to ensure that the
infinite differentiability condition is satisfied. For example, the
function $f^{\prime}(x_{j})$ may have a discontinuity of the same
character as the square wave. Then the discontinuity will induce
oscillations, known as the Gibbs phenomenon. The Gibbs phenomenon
influences the accuracy of the FFT not only in the neighborhood of
the point of singularity, but also over the entire computational
domain. Since the Gibbs phenomenon is related to the slow decay of
the Fourier coefficients of the discontinuous function, it is nature
to use smoothing procedures, which attenuate higher order Fourier
coefficients to damp the oscillations \cite{Spectral-Methods-Book,
Spectral-Methods-Book-2,AIAA,WFFT}. A straightforward way is to
multiply each Fourier coefficients by a smoothing factor
$\sigma_{k}$, for instance, the Lanczons smoothing factor, the
raised cosine smoothing factor, or the Fejer smoothing factor, etc
\cite{Spectral-Methods-Book,AIAA,WFFT}.

In the recent work \cite{PLA}, we presented a way to construct
smoothing factors. Firstly, we expand $k$ in Taylor series
\begin{eqnarray}
k &=&\frac{\arcsin [\sin (k\Delta x/2)]}{\Delta x/2}  \notag \\
&=&\frac{1}{\Delta x/2}[\sin (k\Delta x/2)+\frac{1}{6}\sin ^{3}(k\Delta x/2)+%
\frac{3}{40}\sin ^{5}(k\Delta x/2)+\frac{5}{112}\sin ^{7}(k\Delta x/2)+...]
\notag \\
&=&\frac{1}{\Delta x/2}\sum_{n=0}^{\infty }\frac{\Gamma (n/2)\delta
_{0,\Theta (n)}\varepsilon (-1+n)}{\sqrt{\pi }n\Gamma (\frac{n+1}{2})}\sin
^{n}(k\Delta x/2)\text{,}  \label{K-ts}
\end{eqnarray}%
where $\Gamma (n)=\int_{0}^{\infty }t^{n-1}e^{-t}dt$ is the Gamma function, $%
\Theta (n)=Mod[-1+n,2]$ is the Mod function and $\varepsilon (-1+n)$
is the unit step function. Next, in order to refrain the Gibbs
oscillation, we should filter out more high frequency waves, or at
least, damp the strengths of high frequency waves. Therefore, $k$
may take the form of an appropriately truncated Taylor series
expansion of sin$(k\Delta x/2)$. For example, $k$ can take the
following forms:
\begin{equation}
k_{1}\mathbf{=}\frac{\sin (k\Delta x/2)}{\Delta x/2}\text{,}  \label{K1}
\end{equation}%
\begin{equation}
k_{2}\mathbf{=}k_{1}+\frac{\sin ^{3}(k\Delta x/2)/6}{\Delta x/2}\text{,}
\label{K2}
\end{equation}%
\begin{equation}
k_{3}\mathbf{=}k_{2}+\frac{3\sin ^{5}(k\Delta x/2)/40}{\Delta x/2}\text{,}
\label{K3}
\end{equation}%
and
\begin{equation}
k_{4}\mathbf{=}k_{3}+\frac{5\sin ^{7}(k\Delta x/2)/112}{\Delta x/2}\text{,}
\label{K4}
\end{equation}%
and then the calculated spatial derivative is second-order,
fourth-order, sixth-order, and eighth-order in precision,
respectively. It is found that $k_{1}$ is consistent with the one
used in Ref. \cite{Shu-ICASE Report}. Finally, smoothing factor for
$k_{1}$ can be expressed as
\begin{equation}
\sigma _{1}=\frac{k_{1}}{k}=\frac{\sin (j\pi /N_{x})}{j\pi /N_{x}}\text{, }%
j=-N_{x}/2,...,N_{x}/2\text{,}
\end{equation}%
and the ones for $k_{2}$, $k_{3}$, and $k_{4}$ can be formulated in a
similar way.

As reported in our recent work \cite{PLA}, the lower-order smoothing
factors, such as $\sigma_{1}$ and $\sigma_{2}$, are much more
effective to damp the strengthens of high frequency waves and may
result in excessively smeared approximations, which are unfaithful
representations of the truth physics. On the other hand, the
higher-order smoothing factors, such as $\sigma_{3}$ and
$\sigma_{4}$, can reserve more higher frequency waves but may not
damp the Gibbs phenomenon when the discontinuities are strong
enough, then cause numerical instability. This is especially true
for the case with shock waves and/or discontinuities. The smoothing
factors should survive the dilemma of stability versus accuracy. In
other words, they should be minimal but make the evolution stable.
In the present study, we focus on the liquid-vapor system without
shock waves and strong discontinuities. Therefore, the FFT scheme
with higher-order smoothing factor, $\sigma_{4}$, is used throughout
our simulations.

\begin{figure}[tbp]
\center {
\epsfig{file=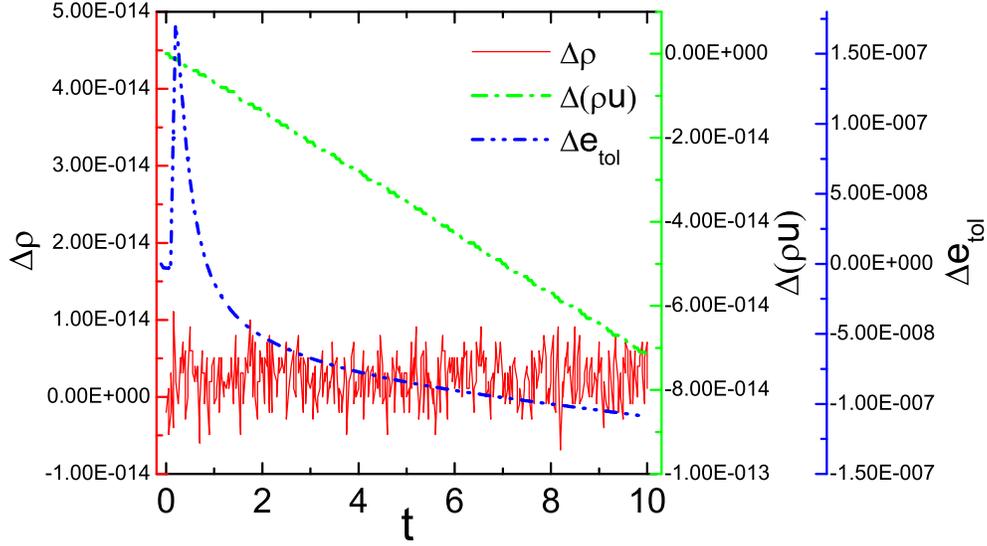,bbllx=9pt,bblly=0pt,bburx=572pt,bbury=362pt,
width=0.8\textwidth,clip=}} \caption{(Color online) Variations of
density $\Delta \rho(t)$, momentum $\Delta (\rho \mathbf{u})(t)$,
and total energy $\Delta e_{tol}(t)$ for the phase-separating system
described in Fig. 1.}
\end{figure}
\begin{figure}[tbp]
\center {
\epsfig{file=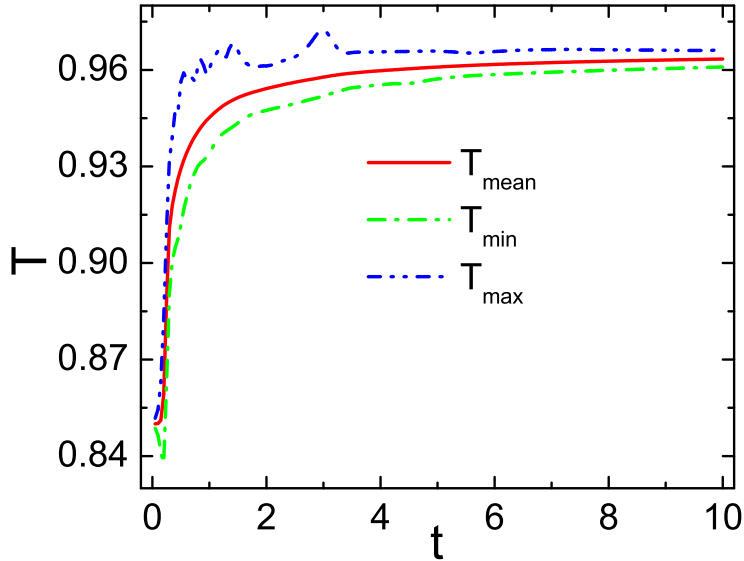,bbllx=3pt,bblly=5pt,bburx=569pt,bbury=445pt,
width=0.6\textwidth,clip=}} \caption{(Color online) Temperatures
versus time $t$ for the FFT-TLB scheme.}
\end{figure}

For comparisons, we verify the proposed FFT algorithm with the same
problem described in Fig. 1 and display variations of density
$\Delta \rho(t)$, momentum $\Delta (\rho \mathbf{u})(t)$, and total
energy density $\Delta e_{tol}(t)$ in Fig. 2, respectively. It is
observed that, when the FFT scheme with $\sigma_{4}$ is adopted,
variations of density and momentum nearly decrease to machine
accuracy. For $\Delta e_{tol}(t)$, it oscillates at the beginning
then goes to nearly a constant. Behaviors of $\Delta e_{tol}(t)$ can
be interpreted as follows. At the beginning of phase separation, the
fluids spontaneously separate into small regions with higher and
lower densities, and more liquid-vapor interfaces appear.
Subsequently, spatial discretization errors induced by the
interfaces (density gradients) arrive at their maxima, accounting
for the initial oscillations. As time evolves further, under the
action of surface tension, the total liquid-vapor interface length
decreases owing to the coalesce of small domains, then the
discretization errors, together with the amplitude of $\Delta
e_{tol}(t)$ decrease.

After about $10^{6}$ time steps, the maximum derivation of $e_{tol}$
is only about $1.5\times 10^{-7}$, indicating that the FFT scheme
has more advantage to guarantee energy conservation. Furthermore, we
find that $\Delta e_{tol}(t)$ decreases with decreasing the initial
random density $\Delta$. When $\Delta$ decreases to $0.001$, the
maximum of $\Delta e_{tol}(t)$ will further decreases to $3\times
10^{-8}$ (not shown here). Numerically, this is owing to the smaller
density gradients in the interface regions as $\Delta$ decreases
that reduce the spatial discretization errors. Actually,
$\Delta=0.001$ ($0.1\%$ of the initial density) is enough to
generate phase separation and is more appropriate. When $\Delta$ is
large, or the initial temperature is far below the critical one, the
initial state of the system is very far from the equilibrium and we
may encounter large values of the fluid velocity in the early stage
of simulations. Since the initial values of the velocity is zero
everywhere, this process is responsible for a strong decrease of the
local temperature (see Fig. 3).

Another interesting phenomenon is illustrated in Fig. 3. The mean
and maximum temperatures rise sharply at the initial period of phase
separation, while the minimum temperature decrease significantly at
first and rise rapidly at later times. The difference between the
minimum and the maximum temperatures $\Delta T =T_{\max}-T_{\min}$
arrives at its maximum at about $t=0.25$. After that, $\Delta T$
decreases with time, and goes to a constant value (nearly vanishing)
when $t>8$. The reasons for behaviors of temperatures are that: at
the initial stage, the potential energy $-9/8\rho^2$, a part of the
free energy, is high, so the system will relax. During phase
separation, part of the potential energy transforms into the kinetic
energy, namely latent heat is locally released and conducts to the
entire region. This is the main reason why temperatures are rising
during simulations and the main difference from isothermal case,
where latent heat is extracted from the system by fixing the
temperature in all lattice nodes. Besides, viscous dissipation is
another mechanism of heat generation. More preciously, heat is
dissipated locally due to the friction between fluid flows when the
fluid velocity is different from zero. After phase separation,
interfaces forms and the fluid velocities go to zero everywhere.
Then the kinetic energy transforms into thermal energy totally.

In our recent work, the FFT-TLB multiphase model has been validated
successfully by two sets of typical benchmarks \cite{PLA}.
Simulation results demonstrated that the FFT-TLB model can capture
both qualitatively and quantitatively the interface properties in
accord with the VDW theory. Besides that, with the new model,
spurious velocities near the liquid-vapor interface are
significantly reduced, and, as a result, phase diagrams of the
liquid-vapor system obtained from simulations are more consistent
with that from theoretical calculations.

\section{Simulation results, rheological and morphological characterizations}

When a system is suddenly quenched into the two-phase region, the
original single phase becomes unstable, then phase separation occurs
through the formation and the subsequent growth of domains.
Eventually, the system arrives at a new equilibrium state. In the
past few decades, this phenomenon has been extensively studied
\cite{Succi-Book,shanchen,Yeomans,Sofonea-multiphase-pre-2004,adv-phy-1994,
Onuki-book,onuki,Chensy-AA=0.67-PRB-1993,Yeomans-PRL-1995,Chensy-PRL-1993,Chensy-jsp-1995,Yeomans-review},
by theoretical derivations, experiments, and numerical simulations.
Among others, the most significant finding is the domain growth law,
which states that, at late times, the characteristic domain size
$R(t)$ grows as a power with time $t$, $R(t)\sim t^{\alpha }$. The
value of exponent $\alpha $ is believed to be universal, depending
only on the growth mechanism, and has been well known in isothermal
system, $\alpha =1/2$ and $2/3$ for high and lower viscosities,
respectively \cite
{Sofonea-multiphase-pre-2004,Chensy-AA=0.67-PRB-1993,Yeomans-PRL-1995,Yeomans-review}
. However, behaviors of phase separation with temperature field are
far from clear. The aim of this section is to clarify effects of
temperature dynamics on both the rheological and morphological
behaviors of phase separation.

\subsection{Patterns for isothermal and thermal cases}

\begin{figure}[tbp]
\center {
\epsfig{file=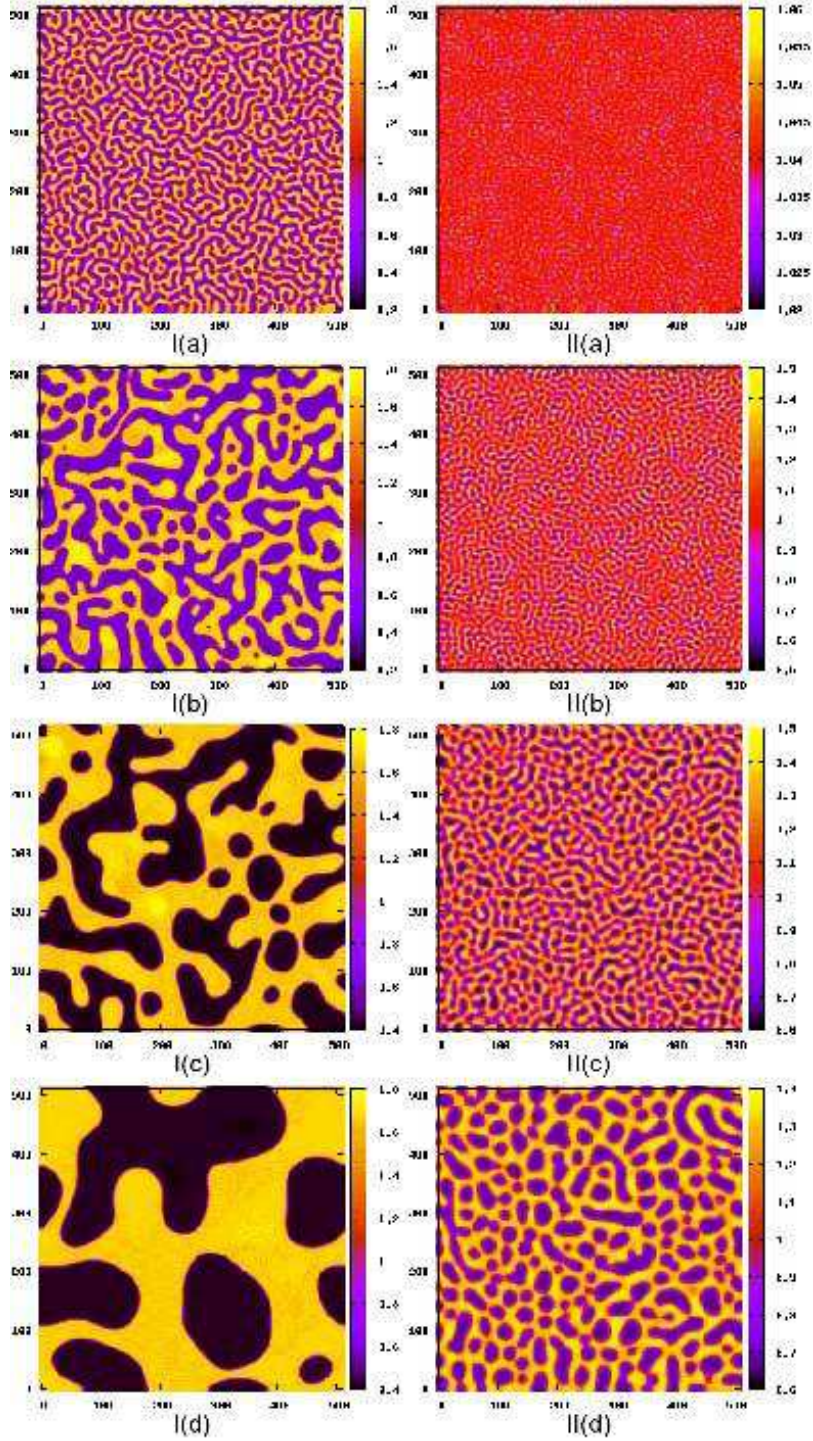,bbllx=0pt,bblly=0pt,bburx=395pt,bbury=711pt,
width=0.65\textwidth,clip=}} \caption{(Color online) Snapshots in
two processes of phase separations. The temperature is fixed at
$T=0.9$ in process I. See Figs. I(a)-I(d). The initial temperature
in process II is $T=0.9$. See Figs. II(a)-II(d). The relaxation time
is fixed at $\protect\tau =10^{-3}$ in the two processes. The time
$t=0.4$, $1.0$, $2.5$, and $8.0$ in (a), (b), (c), and (d),
respectively. The lattice size here is $512\times 512$.}
\end{figure}

Simulations for isothermal and thermal phase separations are
performed on lattices with $N_{x}\times N_{y}=512\times 512$ nodes.
PBC are imposed on both directions. Here, we only consider symmetric
mixtures, namely we set liquid:vapor mass fractions to 1:1, for
which at late times these domains will form a bicontinuous structure
with sharp interfaces \cite{bicontious-prl-1999,bicontious-03-pre}.
Therefore, the initial conditions are set as follows:
\begin{equation}
(\rho, u_{x}, u_{y}, T)=(1.042+\Delta, 0.0, 0.0, 0.9)\text{,}
\label{initial}
\end{equation}
where $1.042$ is the mean density of liquid and vapor at $T=0.9$,
and $\Delta $ is a random density noise with an amplitude of
$0.001$. Parameters are set to be $\tau =10^{-3}$, $\Delta t
=10^{-5}$, $K=5\times 10^{-6}$, $\Delta x=\Delta y=1/256$, and
others are unchanged. Density distribution patterns at representative times $%
t=0.4$, $1.0$, $2.5$, and $8.0$ are shown in Fig. 4 for isothermal
case (see Figs. 4I(a)-I(d)) and thermal case (see Figs.
4II(a)-II(d)). For the isothermal case, after about $25000$ time
steps, the fluid has begun to separate spontaneously into small
regions with higher and lower densities. As time evolves, the small
domains merge with each other and larger domains appear under the
action of surface tension at $t=0.4$. From patterns at $t=0.4$,
$1.0$, and $2.5$, as excepted, higher and lower densities domains
evolve in an equal way, leading to an interwoven bicontinuous
pattern. The growth of domains continues at $t=8.0$, and,
eventually, the system will reach a completely separated state for a
large enough time.

Compared with configurations in the isothermal case, several
distinctive differences can be found in the thermal case: (i) the
average size of domains in each case tends to increase in an effort
to decrease the interfacial energy, while at the same moment, in the
isothermal case, it is bigger than its counterpart, which
demonstrates that the domains grow faster in this case; (ii) for the
isothermal case, interfaces between vapor and liquid are much
clearer, which shows that the interfaces in this case are much
narrower; (iii) density difference between the maximum and minimum
densities $\Delta\rho =\rho_{\max}-\rho_{\min}$ in the isothermal
case is much larger than the one in thermal case, indicating that
phase separation in this case is deeper; (iv) contrary to
interpenetrating bicontinuous structures formed in the isothermal
case, isolated and nearly circle vapor droplets suspending in the
liquid phase are appeared in thermal case. These differences are
interesting and meaningful. In the following subsections, we will
analysis these differences with the help of rheological description
and Minkowski functionals.

\subsection{Rheological characterization}

In order to further quantify the results shown in Fig. 4, time
evolution of the circularly averaged structure factor $S(k,t)$ is
employed, which is defined as the Fourier transform of the
density-density correlation function. For a discrete system, it can
be stated as
\begin{equation}
S(\mathbf{k},t)=\left\vert \sum_{\mathbf{x}}[\rho (\mathbf{x},t)-\bar{\rho}%
(t)]e^{\textbf{i}\mathbf{k\cdot x}}\right\vert /N \text{,}
\label{strf}
\end{equation}
where $\mathbf{k}=(2\pi /N)(m \hat{i} +n \hat{j})$ is the wave
vector in the reciprocal space with $m=1,2,...N_{x}$,
$n=1,2,...N_{y}$. $S(\mathbf{k},t)$ is further smoothed by averaging
over an entire shell in $\mathbf{k}$ space to obtain the circularly
averaged structure factor
\begin{equation}
S(k,t)=\sum_{k}S(\mathbf{k},t)/\sum_{k}1\text{.}  \label{cir_strf}
\end{equation}

\begin{figure}[tbp]
\center {
\epsfig{file=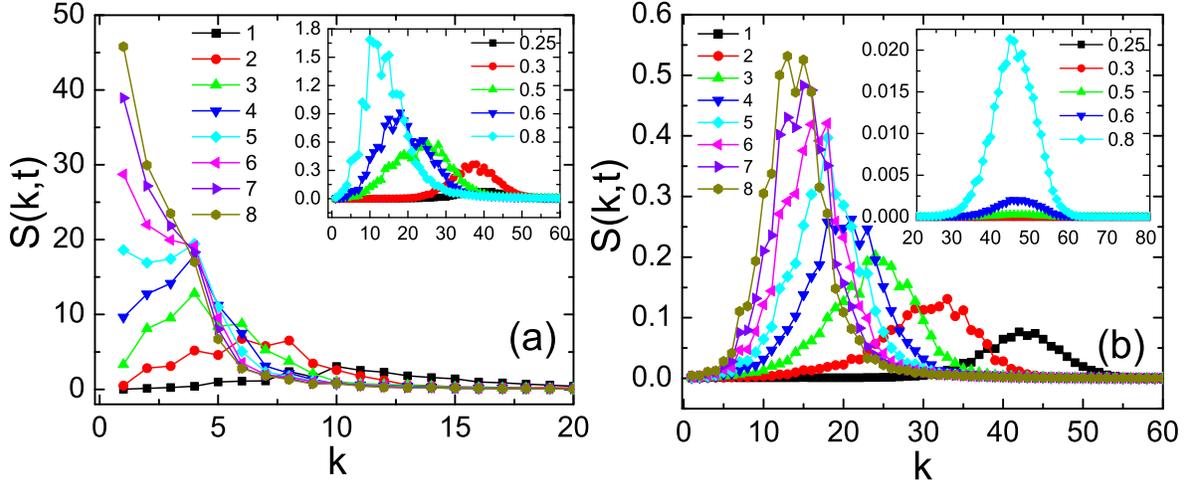,bbllx=1pt,bblly=9pt,bburx=533pt,bbury=230pt,
width=0.95\textwidth,clip=}} \caption{(Color online) Spherically
averaged structure factor $S(k,t)$ versus wave number $k$ for the
procedures shown in Fig. 4. Figure (a) is for the isothermal case
and Fig. (b) is for the thermal case. In each figure, $S(k,t)$ at
times $t=0.25$, $0.3$, $0.5$, $0.6$, and $0.8$ are shown in the
inset, while those at times $t=1.0$, $2.0$, $3.0$, $4.0$, $5.0$, $
6.0$, $7.0$, and $8.0$ are shown in the main frame.}
\end{figure}

In Fig. 5, we present the time evolutions of $S(k,t)$ for isothermal
case in (a) and thermal case in (b), respectively. All curves in
Fig. 5 can be roughly divided into two different time regimes: the
SD stage and the domain growth (DG) stage. From Fig. 5(a), at early
times, such as $t=0.25$ and $0.3$, we observe that the peak in
$S(k,t)$ increases in height without the position of the peak
changing in time. This behavior is indicative of the initial
sharpening of domains, without detectable phase separation taking
place. In the second stage, the peak of $S(k,t)$ increases in height
and shifts to smaller wave number, indicating the coarsening of
domains. At $t=0.5$, $0.6$, and $0.8$, we observe the appearance of
a second peak in $S(k,t)$, which merges with the main peak later on.
This behavior manifests that there is more than one typical domain
size at that moment. From $t=6.0$ onwards, the peak seems to stop
drifting to the left but only oscillates in amplitude, which means
that the finite size effects are pronounced.

Similar results can also be found in the thermal case. Nevertheless,
careful comparisons of these two cases will show you some
distinctions: (i) the first stage continues up to $t=0.8$, which is
longer than the isothermal case. The existence of temperature field
significantly decelerate the speed of domain formation, an effect
which has also been seen in Fig. 4; (ii) over the period from
$t=4.0$ to $t=6.0$, the peak of $S(k,t)$ only varies in height but
very little in wave number. This phenomenon is usually observed at
the initial stage of phase separation, leading us to think that the
system has steered to a new SD stage before reaching the finial late
time stage. Essentially, during this stage, the dynamics is mainly
making the interfaces thinner while the average domain sizes barely
change; (iii) at the same time, the peak of $S(k,t)$ in isothermal
case is much larger than the one in thermal case, but the
corresponding wave number is much smaller, which demonstrate that
both the density difference between the two phases and the
characteristic domain size are much larger in the isothermal case.
These results agree with the information obtained from Fig. 4.

\begin{figure}[tbp]
\center {
\epsfig{file=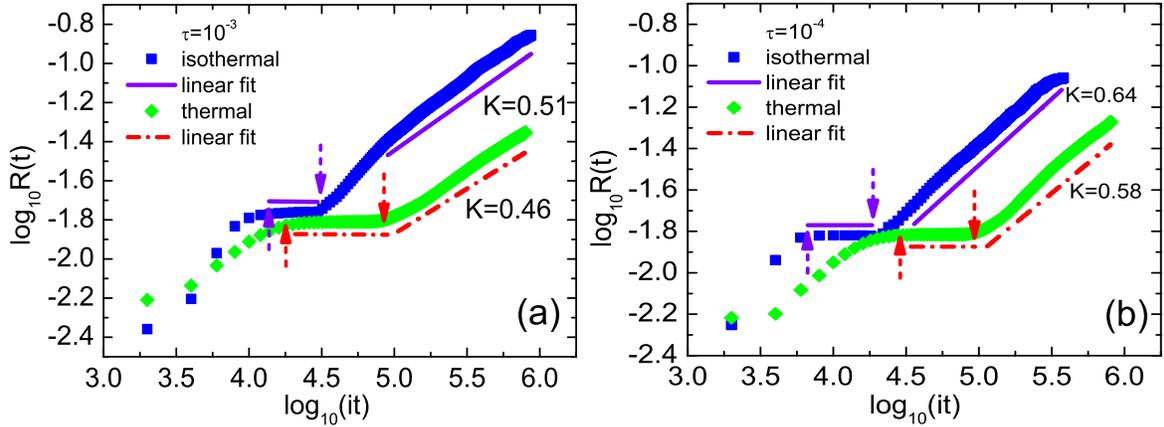,bbllx=0pt,bblly=0pt,bburx=532pt,bbury=204pt,
width=0.95\textwidth,clip=}} \caption{(Color online) Domain growths
in isothermal and thermal systems with $\protect\tau =10^{-3}$ in
(a) and $\protect\tau =10^{-4}$ in (b). The squares and diamonds are
for results from FFT-TLB simulations. Lines are shown in each plot
to guide the eyes. }
\end{figure}

Next, the characteristic domain size $R(t)$ is used to further
describe the kinetic process quantitatively. $R(t)$ is derived from
the inverse first moment of $S(k,t)$,
\begin{equation}
R(t)=2\pi \sum_{k}S(k,t)/\sum_{k}kS(k,t)\text{.}
\end{equation}
In Fig. 6, we show the growth of $R(t)$ versus iterations for, $\tau
=10^{-3}$ in (a), and $\tau =10^{-4}$ in (b), in a log-log scale. In
each figure, the top and bottom scatter symbols correspond to the
simulation results for the isothermal and thermal cases,
respectively. Straight lines in each plot are linear fits of the
simulation results. Discarding both the early time transient regime
and the very late time regime, where finite-size effects are
pronounced, we find, for isothermal case, the behaviors of $R(t)$
during the DG stage are $R(t)\sim t^{0.52}$ for $\tau =10^{-3}$, and
$R(t)\sim t^{0.64}$ for $\tau =10^{-4}$. These results are in good
agreement with the generally accepted theoretical predictions of $R(t)\sim t^{1/2}$ and $%
R(t)\sim t^{2/3}$ at high and low viscosities by the Allen-Cahn
theory \cite{Allen-Cahn-aa=0.5=1979} and LB models
\cite{Sofonea-multiphase-pre-2004,Chensy-AA=0.67-PRB-1993,Yeomans-PRL-1995,Yeomans-review}%
. But for the thermal case, the growth exponents decrease to $0.46$
for $\tau =10^{-3}$, and $0.58$ for $\tau =10^{-4}$, respectively.
This can be regarded as another proof for our conclusion, which
states that domains grow faster at lower temperature. In addition to
the above differences, another piece of information also deserves
our attention. For the isothermal case with $\tau =10^{-3}$, the SD
stage lasts about for $25000$ time steps (see the upper horizontal
solid line in Fig. 6(a)). Nevertheless, for the thermal case, it
lasts for $80000$ time steps (see the lower horizontal dash dot line
in Fig. 6(b)). Similar results can also be found in the case with
$\tau=10^{-4}$. These findings suggest that, compared to the
isothermal case, the SD stage is significantly prolonged by the
existence of temperature field. In the following part, the
morphological functionals are used to verify similarities and
differences between the two cases, and the corresponding physical
interpretations are given.

\subsection{Morphological characterization and physical interpretations}

\subsubsection{Similarities and differences}

\begin{figure}[tbp]
\center {
\epsfig{file=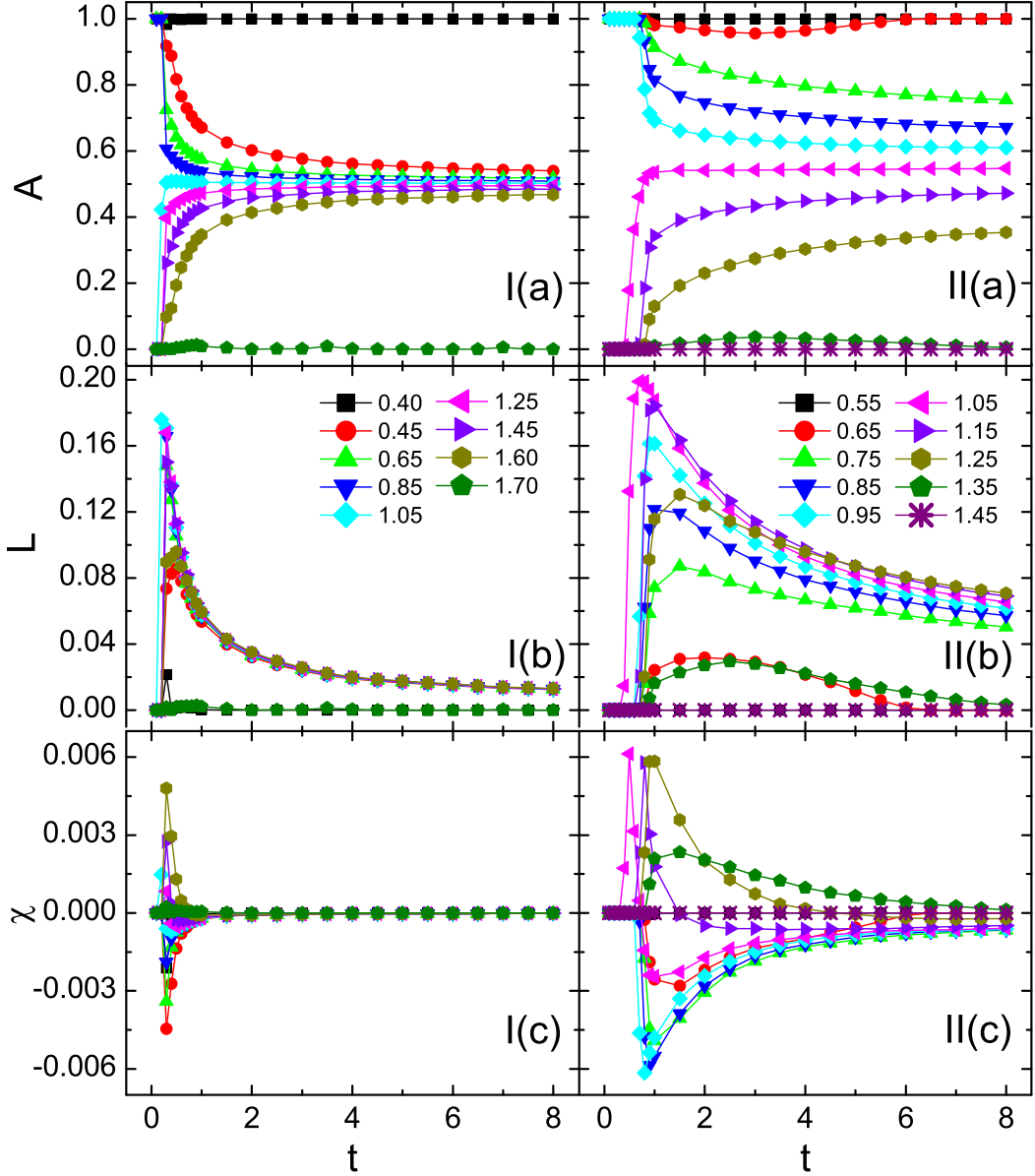,bbllx=12pt,bblly=5pt,bburx=573pt,bbury=648pt,
width=0.85\textwidth,clip=}} \caption{(Color online) Minkowski
measures for the procedures shown in Fig. 4. The left column is for
the isothermal case and the right column is for the thermal case.}
\end{figure}

To perform Minkowski functionals analysis for the density map, we
choose a density threshold $\rho_{th}$ and pixelize the map into
high density regions (with $\rho>\rho_{th}$) and low density regions
(with $\rho<\rho_{th}$). Figure 7 shows the time evolutions of
Minkowski measures for the procedures shown in Fig. 4. From Fig.
7I(a) for the isothermal case, we see that, when $\rho_{th}=0.40$,
the white area fraction $A$ keeps nearly $1.0$ during the whole
procedure shown here, which means no local density is lower than
$0.40$ in the system up to $t=8.0$. However, when the threshold
increases to $1.70$, the white area fraction $A$ keeps nearly zero
during the whole process. Thus, no local density is higher than
$1.70$ in the system. As a direct consequence of SD, $A$ increases
with time $t$ when $\rho_{th}>1.0$, while it decreases when
$\rho_{th}<1.0$. Consequently, most curves (except for the uppermost
curve for $\rho_{th}=0.40$ and the lowermost curve for
$\rho_{th}=1.70$) toward the horizontal central line from about
$t=0.25$. Afterwards, at the DG stage, the curve for
$\rho_{th}=1.05$ (mean density of the system) overlaps with the
horizontal central line and other curves are symmetric to it. The
outer two curves are for the cases with $\rho_{th}=0.45$ (the red
ball) and $\rho_{th}=1.60$ (the dark yellow hexagon), respectively.
We mention that these two densities are just the equilibrium
densities of vapor and liquid at $T=0.9$. Obviously, the outer two
curves mark the reach of the correct equilibrium state. As time
evolves, proportion between the two curves decreases, as a
consequence of the coarsening of the high/low density domains.
Moreover, the proportion between any two curves can be conveniently
obtained from Fig. 7I(a).

Now we go to the second and the third Minkowski measures, the
boundary length $L$ and Euler characteristic $\chi$. As shown in
Fig. 7I(b), for each case, $L$ increases sharply to its maximum at
about $t=0.25$, then decreases slowly. The first increase and the
subsequent decrease in $L$ are due to the appearance of the
liquid-vapor interface during the SD stage and the following
decrease in interfacial area during the DG stage, respectively. At
the SD stage, when $\rho_{th}<1.0$, $\chi$ decreases to be evidently
less than zero, which indicates that the number of domains with
$\rho<\rho_{th}$ increases. On the contrary, when $\rho_{th}>1.0$,
$\chi$ increases to be evidently larger than zero, which indicates
that the number of domains with $\rho>\rho_{th}$ increases. These
results show that the phase separation process is in progress. We
mention that, at about $t=0.25$, the case with $\rho_{th}=0.45$ has
the minimum Euler characteristic and the case with $\rho_{th}=1.60$
has the maximum one, but the two cases get the minimum boundary
length $L$. These results indicate the following information: for
the first case, many scattered black domains with $\rho_{th}<0.45$
appear in the high density background with $\rho_{th}>0.45$, while
for the second case, the high density domains with $\rho_{th}>1.60$
are scattered in the low density background with $\rho_{th}<1.60$.
These domains are so small that the total boundary length is nearly
zero. From Figs. 4I(c)-I(d), we observe that the density maps show
highly connected structures with nearly equal and very small numbers
of black and white domains. Hence, the Euler characteristic $\chi$
keeps close to zero in the DG stage (see Fig. 7I(c)).

From Figs. 7II(a)-II(c), for thermal case, one can also distinguish
two different stages. At early times ($t<0.8$), due to the growth of
density fluctuations and the build up of interfaces, the density
area fraction $A$ belonging to the liquid phase increases, while the
one belonging to the vapor phase decreases. The changes also result
in the increase in boundary length $L$ (see Fig. 7II(b)). The
appearance of liquid-vapor interfaces has an additional effect. They
separate the system with disconnected minority domains. As a result,
the absolute value of Euler characteristic $\chi$ increases in the
SD stage. In contrast to the first stage, as a direct consequence of
the coalescence of relatively small domains, the characteristic
length scale increases but the number of domains decreases,
therefore, the DG stage ($t>0.8$) is characterized by the decrease
in $L$ and $\chi$.

In the end of the second stage, an interesting phenomenon occurs.
There are two small proportions for $\rho<0.65$ and $\rho>1.35$
during the second stage and reach their maxima at about $t=3.0$ (see
Fig. 7II(a)), but are gradually diminishing afterwards. This
phenomenon shows that a recombination process is taking place owing
to the increasing temperature that interrupts the original process
and forces the system evolves to a new equilibrium state decided by
the variable temperature.

Compared with figures shown in Fig. 7, main differences between
these two cases are analysed and listed as follows: (i) for
isothermal case, the domain with a density between $[0.45,1.60]$
only accounts for $20\%$ at $t=4$, and decreases further with time.
But for thermal case, the domain with a density between
$[0.75,1.25]$ reaches to $50\%$ of the whole domain at $t=8$. The
difference demonstrates that the separation depth in thermal case is
much shallower, while the interface width is much wider, which can
be clearly seen in Fig. 7II(b); (ii) the maxima of $L$ and $\chi$
can be used to mark the transition from the SD stage to DG stage
\cite{sofonea-PRE-1997}. The transition time for isothermal case is
about $0.25$, but for thermal case, it increases to about $0.8$. The
result further confirms our conclusion that: phase separation occurs
faster in the isothermal case. From another point of view, this
conclusion can also be obtained from the $A(t)$ curve. For most
cases, after the initial quick changing period, the changing of $A$
with time $t$ shows a slowing down. The slope of the $A(t)$ curve
corresponds approximately to the speed of phase separation. For the
same density threshold, the slopes of $A(t)$ curves in the two cases
are quite different. For example, when $\rho_{th}=0.65$, the $A(t)$
curve decreases sharply in isothermal case, while for the thermal
case, it decreases much more slowly. So we can say that, in
isothermal case, the process of phase separation is much faster;
(iii) connectivity of patterns in isothermal case is much better
than that in thermal case. This feature can be achieved from the
evolution of $\chi$. $\chi$ decreases enormously and almost vanishes
at about $t=0.5$ in the isothermal case, but it is negative for the
vapor structure until about $t=6.0$ in the thermal case, which is
consistent with density patterns in Fig. 4 (see Fig. 4II(d)).

\subsubsection{Physical interpretations of the prolonged SD stage and the
lower growth exponent in thermal case: effects of temperature and
viscosity}

In Sec. IV, we find, compared to isothermal case, the SD stage is
significantly prolonged and the growth exponent is lowered in
thermal case. In this subsection, effects of temperature and
viscosity are investigated to provide proper interpretations.

\begin{figure}[tbp]
\center {
\epsfig{file=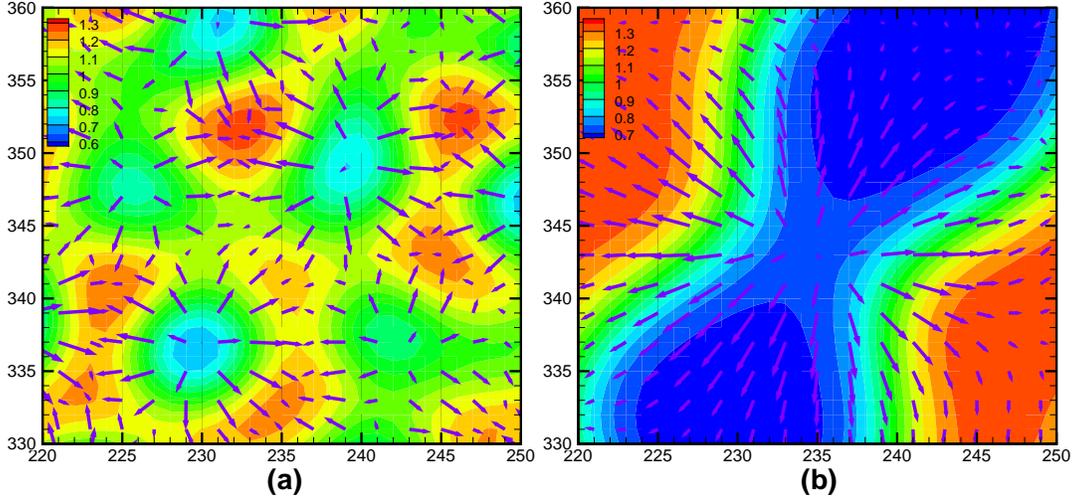,bbllx=74pt,bblly=547pt,bburx=551pt,bbury=773pt,
width=0.87\textwidth,clip=}} \caption{(Color online) Portion of the
density and temperature gradient
distributions at $t=0.8$ in (a) and $t=7.0$ in (b) for the thermal case with $%
\protect\tau =10^{-3}$. The lengths of the temperature gradient
vectors are magnified by $400$ times in (a) and $4000$ times in
(b).}
\end{figure}

Firstly, in Fig. 8, we display the density and the corresponding
temperature gradient distributions at two representative times for
the thermal case with $\tau =10^{-3}$. To illustrate the structure
of the temperature gradient fields clearly, the lengths of the
vectors are multiplied by $400$ in (a) and $4000$ in (b),
respectively. As shown in Fig. 8(a), many tiny droplets and bubbles
appear in the system, and the temperature gradient vectors are
toward the droplets. Thus, the local temperatures within droplets
are slightly higher than the mean temperature of the system, while
the local temperatures within bubbles are slightly lower than the
mean temperature. With the separating process, the local
temperatures in the two phases deviate more from the mean
temperature and an overshoot phenomenon is observed. This procedure
continues up to an extent, after which the local phases with high
(low) temperatures partly begin to transform back from liquid
(vapor) to vapor (liquid). In this way, both the local high
temperatures and low temperatures approach the mean temperature, and
the system approaches thermodynamical equilibrium quickly at lower
Pr number ($\Pr =0.1$ for $\tau =10^{-3}$). This process is evident
by Fig. 8(b), where there is no determinate relationship between
temperature gradients and liquid (vapor) domains. Moreover, the
temperature difference between the highest and lowest decreases to
$0.014$. The system approaches to thermodynamical equilibrium so
quickly that the temperature difference becomes so small during the
phase separation process. Therefore, in this case, temperature can
not be regarded as an ideal physical quantity to describe this
process.

\begin{figure}[tbp]
\center {
\epsfig{file=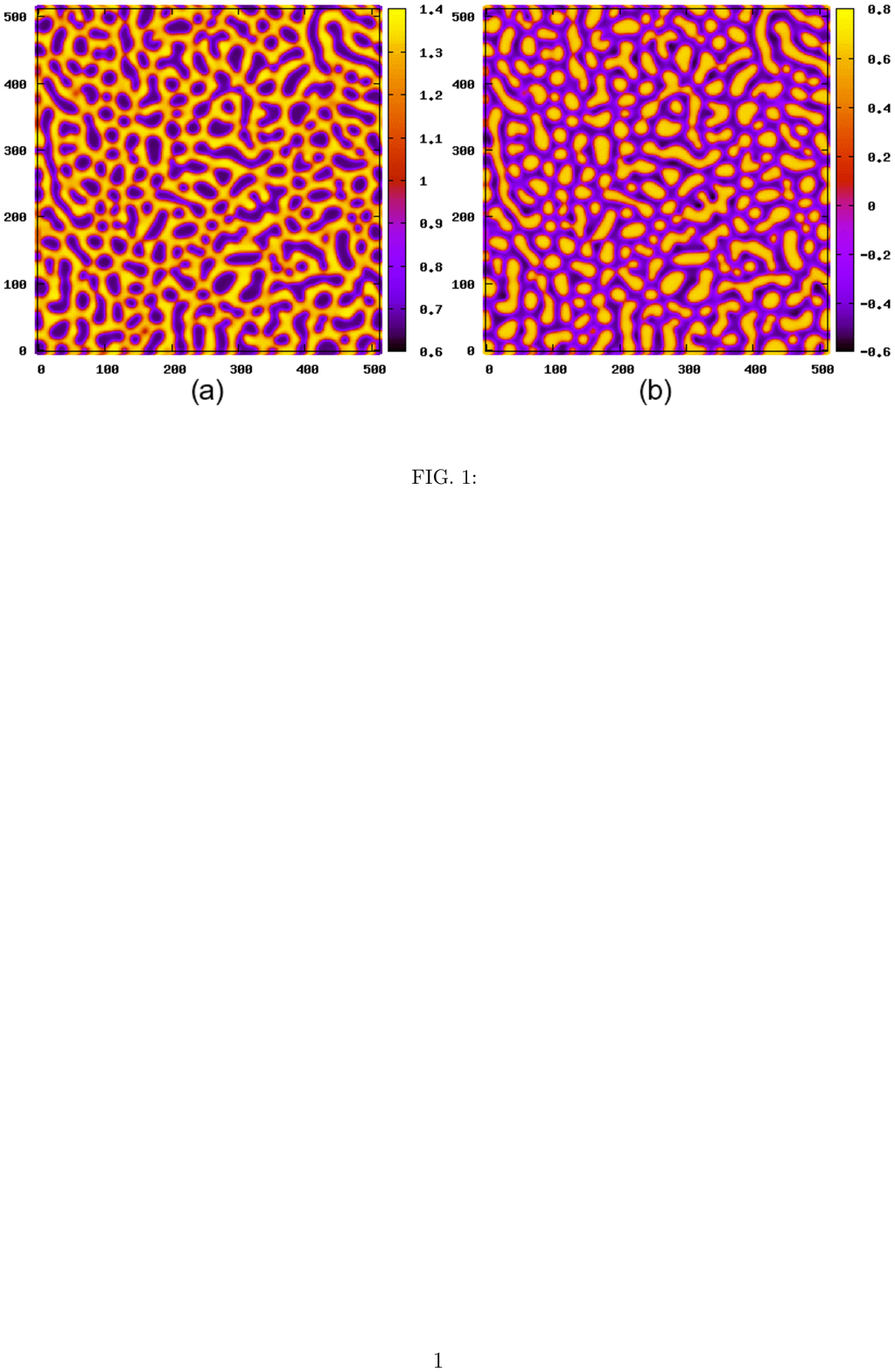,bbllx=72pt,bblly=553pt,bburx=545pt,bbury=771pt,
width=0.9\textwidth,clip=}} \caption{(Color online) Density (a) and
enthalpy (b) distributions at $t=6.0$ for the thermal case with
$\protect\tau =10^{-3}$.}
\end{figure}

In another way, we employ enthalpy and latent heat to describe this
process, and the enthalpy is defined by
\begin{equation}
h=\epsilon +P/\rho\text{,}  \label{han}
\end{equation}
where $\epsilon =\rho T-9/8\rho^{2}+K\left\vert \nabla \rho
\right\vert^{2}/2 $ is the internal energy density including the gradient
contribution. The difference of enthalpy between two states determines the
latent heat $L_{h}$
\begin{equation}
L_{h}=h_{2}-h_{1}.  \label{Latent heat}
\end{equation}%
Figure 9 shows the density and enthalpy distributions at $t=6.0$ for
the thermal case with $\tau =10^{-3}$. Comparison of the two figures
illustrates that the enthalpy of vapor is relatively higher than
that of liquid. In order to study the dynamic characteristics of the
pattern, we show the spatial distribution of latent heat
$L_{h}=h_{t=8.0}-h_{t=3.0}$ in Fig. 10(a) and density change
$\Delta\rho=\rho_{t=8.0}-\rho_{t=3.0}$ between these two states in
Fig. 10(b). Careful observations between these two figures suggest
that when the latent heat $L_{h}$ is positive, the corresponding
density difference $\Delta\rho$ is negative. Droplets (Bubbles)
absorb latent heat and evaporation occur simultaneously.
Subsequently, the density decreases. A negative $L_{h}$ corresponds
to an increase of density, then the droplets (bubbles) have a
coagulation trend. It should be noted that, owing to the
transformation of potential energy into thermal energy, the total
latent heat is released during the whole process. The released heat
conducts over the entire region rapidly at low Pr number, and
increases the mean temperature of the system (see Fig. 11). While in
isothermal system, latent heat is extracted from this system by
fixing the temperature in all lattice nodes.

\begin{figure}[tbp]
\center{
\epsfig{file=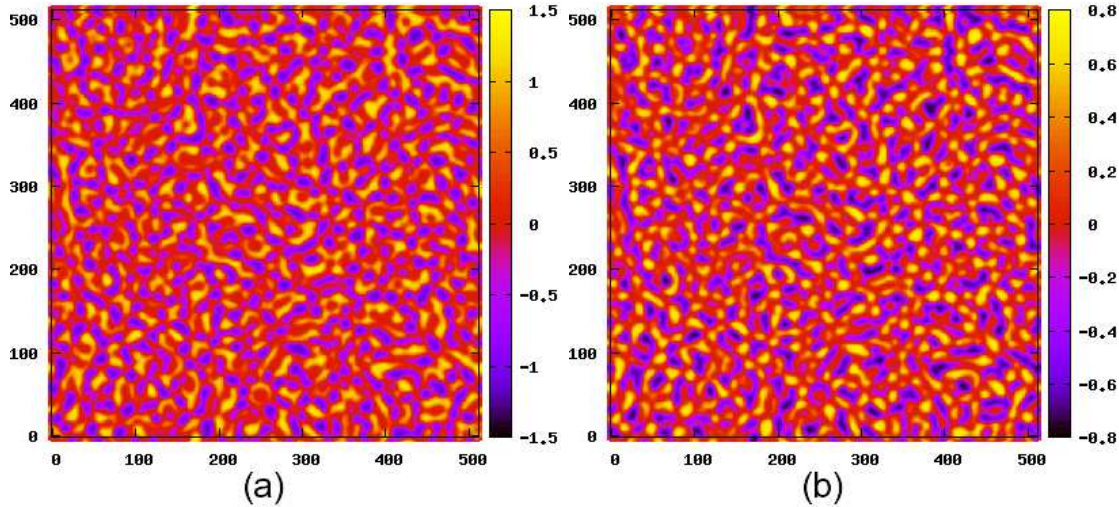,bbllx=72pt,bblly=553pt,bburx=545pt,bbury=771pt,
width=0.9\textwidth,clip=}}
\caption{(Color online) Distribution of latent heat (a) $%
L_{h}=h_{t=8.0}-h_{t=3.0}$ and the corresponding distribution of
density change (b) $\Delta\protect\rho
=\protect\rho_{t=8.0}-\protect\rho_{t=3.0}$ for the thermal case
with $\protect\tau =10^{-3}$.}
\end{figure}
\begin{figure}[tbp]
\center {
\epsfig{file=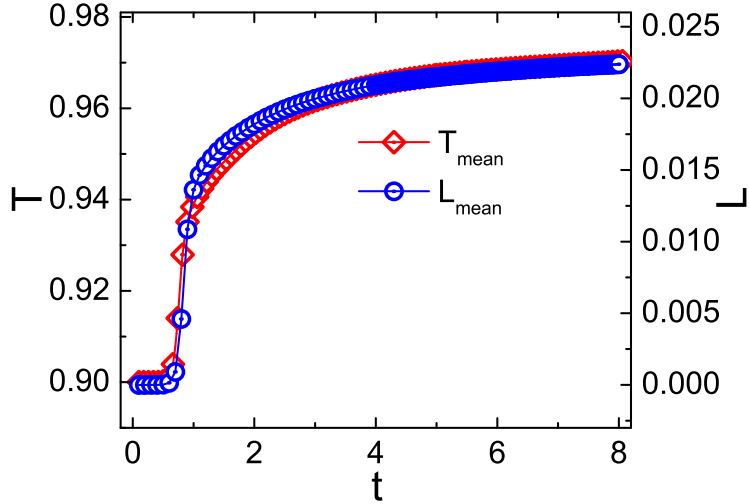,bbllx=3pt,bblly=3pt,bburx=562pt,bbury=394pt,
width=0.6\textwidth,clip=}} \caption{(Color online) Time evolutions
of the mean temperature and released latent heat for the procedures
shown in Figs. 4 II(a)-II(d).}
\end{figure}

As well, another piece of information can be obtained from Fig. 11.
The mean temperature scarcely grows at the first stage due to no
detectable phase separation is taking place and no remarkable latent
heat is released. Subsequently, in the next stage, the temperature
rapidly increases to $0.97$ and, later, keeps almost zero growth.
Afterwards, under the almost unchanged temperature, phase separation
evolves in accord with the isothermal case.

\begin{figure}[tbp]
\center {
\epsfig{file=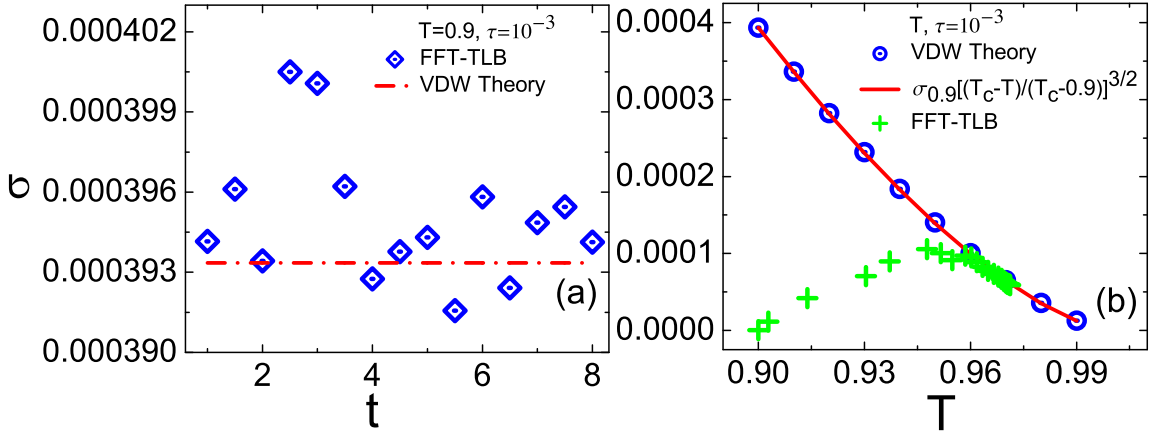,bbllx=8pt,bblly=3pt,bburx=579pt,bbury=225pt,
width=0.95\textwidth,clip=}} \caption{(Color online) Time evolution
of surface tension for the procedures shown in Fig. 4. Figure (a) is
for the isothermal case and Fig. (b) is for the thermal case.}
\end{figure}

So far, we have not discussed in detail the surface tension. For a
planar interface, it can be computed from the following formula
\cite{kk-1,kk-2,kk-3}:
\begin{equation}
\sigma =K\int_{-\infty }^{\infty }(\frac{\partial \rho }{\partial z})^{2}dz%
\text{,}  \label{TiDu}
\end{equation}%
or from the VDW theory \cite
{Interface-2-PRA-1975,Interface-3-PRA-1979,Interface-1-PRE-2008}
\begin{equation}
\sigma =\frac{(aK)^{1/2}}{b^{2}}\sigma ^{\ast}=\frac{(2aK)^{1/2}}{b^{2}}%
\int_{\rho _{g}^{\ast }}^{\rho _{l}^{\ast }}[\Phi ^{\ast }(\rho ^{\ast
})-\Phi ^{\ast }(\rho _{l}^{\ast })]^{1/2}d\rho ^{\ast },  \label{ST}
\end{equation}
where
\begin{equation}
\Phi^{\ast }=\rho ^{\ast }\xi -\rho ^{\ast }T^{\ast }[\ln (1/\rho ^{\ast
}-1)+1]-\rho ^{\ast 2},  \label{fai}
\end{equation}%
\begin{equation}
\xi =T^{\ast }\ln (1/\rho _{s}^{\ast }-1)-\rho _{s}^{\ast }T^{\ast
}/(1-\rho _{s}^{\ast })+2\rho _{s}^{\ast }\text{, }s=v\text{,} l
\label{kexi}
\end{equation}%
$\rho ^{\ast }=\rho b$, $T^{\ast}=bT/a$, with $a=9/8$ and $b=1/3$ in
this model. It is pointed out that Eq. (\ref{ST}) is especially
convenient, since it can be evaluated directly without determining
the density profile. Now, we calculate surface tension with Eq.
(\ref{TiDu}) for both the isothermal and thermal cases from profiles
along the $x$-axis and, at the same time, calculate theoretical
values from Eq. (\ref{ST}). These results are plotted in Fig. 12. It
is shown that, in the isothermal case, after the formation of
liquid-vapor interface, the surface tension keeps nearly a slightly
oscillating constant around the exact value. While in thermal case,
the surface tension is much smaller than the theoretical one before
interfaces are well formed (before the mean temperature reach to
$0.96$). After that, it decreases obviously with the increase of
temperature, and can be verified in the following form:
\begin{equation}
\sigma =\sigma _{0.9}[(T_{c}-T)/(T_{c}-0.9)]^{3/2},
\end{equation}
where $\sigma_{0.9}$ is the surface tension at $T=0.9$. The
increasing temperature lowers the density gradient, as well as the
surface tension, which is the driving force for diffusive growth. As
a result, domains grow more slowly than in the isothermal case.

Essentially, during the whole process, compared to the isothermal
case, two competition mechanisms exist. The first one is heat
generation and conduction mechanism, or temperature rising
mechanism. The release of latent heat results in a rising
temperature and the rising temperature results in a new dependence
of pressure-density. In other words, it leads to a new local
mechanical balance. The second one is the hydrodynamic flow
generation and development mechanism, or liquid-vapor equilibrium
mechanism, decided by viscosity, diffusivity of the fluid, etc. They
compete and influence with each other, deciding the finical
morphology jointly.

\begin{figure}[tbp]
\center {
\epsfig{file=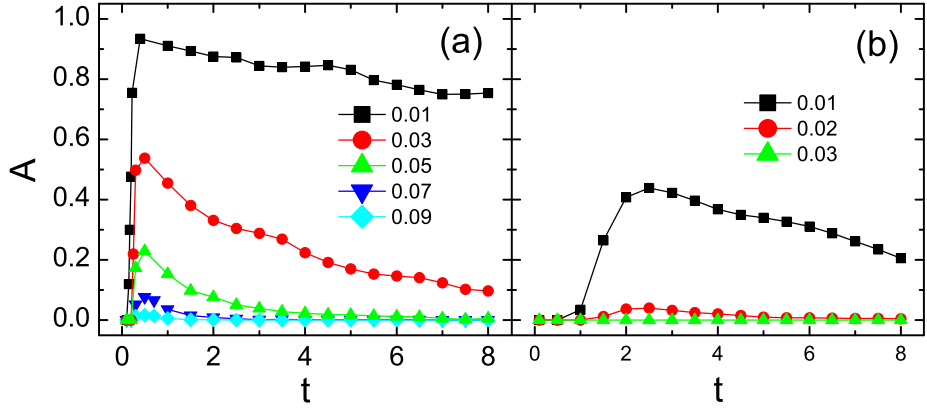,bbllx=4pt,bblly=4pt,bburx=572pt,bbury=259pt,
width=0.75\textwidth,clip=}} \caption{(Color online) High flow
velocity area fraction $A$ versus time $t$ for the procedures shown
in Fig. 4. Figure (a) is for the isothermal process and Fig. (b) is
for the thermal case.}
\end{figure}
\begin{figure}[tbp]
\center {
\epsfig{file=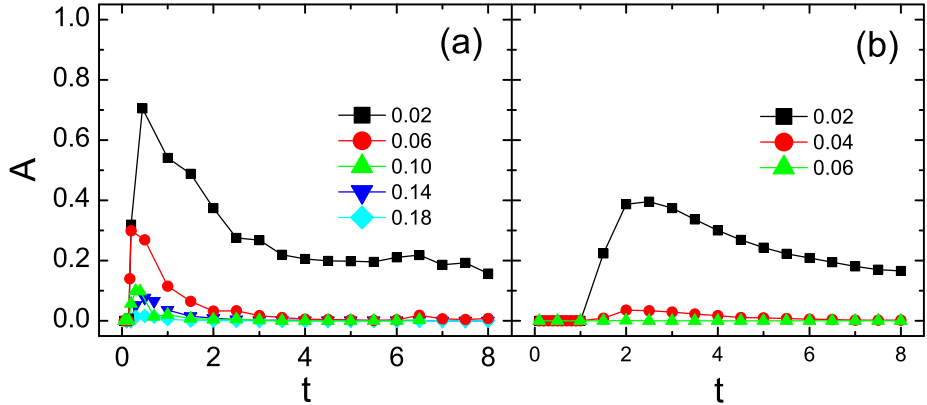,bbllx=4pt,bblly=4pt,bburx=572pt,bbury=259pt,
width=0.75\textwidth,clip=}} \caption{(Color online) High flow
velocity area fraction $A$ versus time $t$ for two phase separation
procedures. Figure (a) is for the isothermal process
and Fig. (b) is for the thermal case. Here, the relaxation time is set to be $\protect%
\tau =10^{-4}$.}
\end{figure}

Besides the temperature effects, we now consider how the
hydrodynamic flows influence both the morphology of the phase
separating system and the growth exponent. Figure 13 shows the time
evolution of the high velocity $\left\vert\mathbf{u}\right\vert $
area fraction for the procedures shown in Fig. 4. From it, also, two
stages can be found, corresponding to the nearly-zero value of white
area fraction for all $\left\vert\mathbf{u}\right\vert_{th}$, the
rapidly increase and the subsequent slowly decrease. For isothermal
case, $\left\vert \mathbf{u}\right\vert\in\lbrack 0,0.09]$, while
for thermal case, $\left\vert \mathbf{u}\right\vert\in\lbrack
0,0.02]$. This implies that velocities are not only damped by
viscosity but also by the rising temperature. The maximum velocity
in the isothermal case can reach to $0.09$ or even higher.
Consequently, in contrast to thermal case, hydrodynamic effects can
not be totally neglected in the isothermal case. The appearance of
larger flow velocities offers more opportunities for coalescence
between domains. Under the action of diffusion and hydrodynamic
flows, a faster DG process is taking place, and a bigger growth
exponent can be observed. Figure 14 shows the same trend of the
hydrodynamic flows, when $\tau$ decreases to $10^{-4}$. Due to the
lower viscosity, velocities are more sufficiently developed. Figures
13 and 14 demonstrate that, in the thermal case, compared to the
thermodynamic and diffusion mechanisms, hydrodynamic flows are less
important than that in the isothermal case and, therefore, can not
be regarded as a dominant factor governing the growth exponent.

\begin{figure}[tbp]
\center {
\epsfig{file=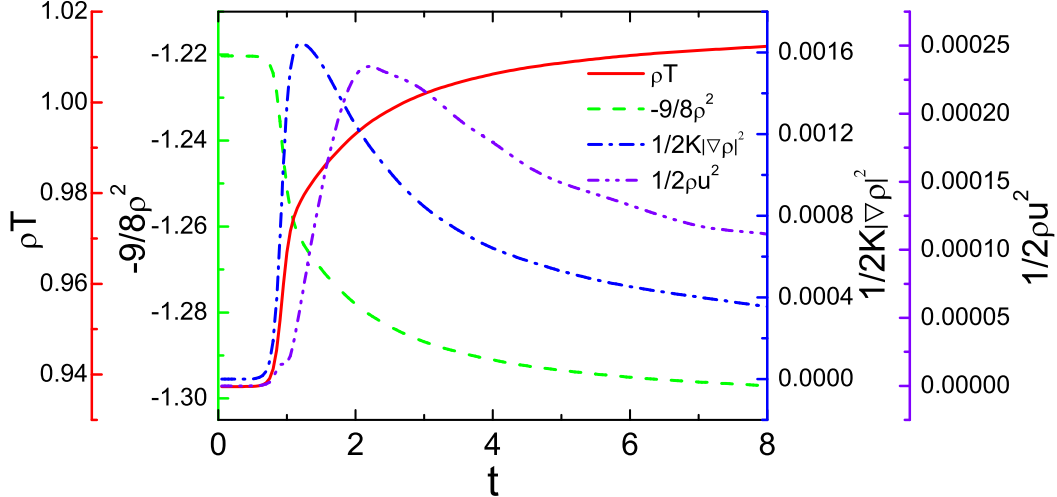,bbllx=3pt,bblly=3pt,bburx=571pt,bbury=284pt,
width=0.85\textwidth,clip=}} \caption{(Color online) Time evolution
of each part of total energy for the procedures shown in Figs.
4II(a)-II(d).}
\end{figure}

In Fig. 15, we display the time evolution of each part of total
energy for the procedures shown in Figs. 4II(a)-II(d), which
presents a clear image about energy evolution during phase
separation. It is found that, $\rho T$ and $-9\rho^{2}/8$ are the
main parts of total energy, and evolve in the opposite way. Kinetic
energy and surface energy are much smaller than the two former ones.
The maximum of $K\left\vert\nabla\rho \right\vert^{2}/2$ corresponds
to the appearance of nuclei and formation of small domains.
Afterwards, it decreases gradually due to the increasing temperature
and the decreasing interfacial area. The macroscopic kinetic energy
$\rho u^{2}/2$ is so small that the viscous dissipation induced by
it can be neglected. Therefore, in thermal case, compared to latent
heat, the effects of kinetic energy on temperature are less
important.

\section{Conclusions and discussions}

Thermal and isothermal symmetric liquid-vapor separations are
simulated via the FFT-TLB method. Structure factor, domain size, and
Minkowski functionals are used to describe the density and velocity
fields and, at the same time, to understand the configurations and
the kinetic processes. Simulations and physical analysis present the
following scenario for the thermal phase separation. When the
separation starts, many tiny droplets and bubbles appear in the
system. The local temperatures within droplets are slightly higher
than the one within bubbles. With separating, neighboring droplets
(bubbles) coalesce and the mean domain size increases. The local
temperatures in the two phases deviate more from the mean
temperature. This procedure continues up to a stage, after which the
local phases with high (low) temperatures partly begin to transform
back from liquid (vapor) to vapor (liquid). In this way, both the
local high temperatures and low temperatures approach the mean
temperature, and the system approaches thermodynamical equilibrium.

Simulation results also indicate that the phase separation in
thermal and isothermal cases can be generally divided into two
stages: the SD stage and the DG stage. Different from the isothermal
case, the SD stage is significantly prolonged, and different
rheological and morphological behaviors are induced by the variable
temperature field in the thermal case. After the transient
procedure, both the thermal and isothermal separations show
power-law scalings in the domain growth; while the exponent for
thermal system is lower than that for isothermal system. With
respect to the density field, the isothermal system presents more
likely bicontinuous configurations with narrower interfaces, while,
the thermal system presents more likely configurations with
scattered bubbles.

Compared with the isothermal case, heat creation, conduction, and
lower interfacial stresses are the main reasons for the differences
in thermal system. Latent heat, is released during the separating
process, which is the main reason for the rising temperature. The
changing of local temperature results in new local mechanical
balance. When the Prandtl number becomes smaller, the system
approaches thermodynamical equilibrium more quickly. The increasing
local temperature has an additional effect. It makes the interfacial
stress lower. This behavior in simulations is quantitatively
verified by the theoretical formula, $\sigma =\sigma
_{0}[(T_{c}-T)/(T_{c}-T_{0})]^{3/2}$, where $T_{c}$ is the critical
temperature and $\sigma_{0}$ is the interfacial stress at a
reference temperature $T_{0}$. Besides thermodynamics, we find that
the local viscosities also influence the morphology of the phase
separating system. For both the isothermal and thermal cases, growth
exponents and local flow velocities are inversely proportional to
the corresponding viscosities. Compared with isothermal case, the
local flow velocities in thermal case not only depend on viscosity
but also temperature. In future studies, we will increase the depth
of separation which the FFT-TLB model can undergo, and investigate
quantitatively how the Prandtl number affects the separation
procedure.

\section*{Acknowledgements}

The authors sincerely thank the anonymous reviewer for her/his
valuable comments and suggestions, and we warmly thank Dr. Victor
Sofonea for many instructive discussions, and also Dr. Qinli Zhang,
and Dr. Bohai Chen for many useful suggestions. AX and GZ
acknowledge support of the Science Foundations of LCP and CAEP
[under Grant Nos. 2009A0102005, 2009B0101012], National Natural
Science Foundation of China [under Grant No. 11075021]. YG and YL
acknowledge support of National Basic Research Program (973 Program)
[under Grant No. 2007CB815105], National Natural Science Foundation
of China [under Grant No. 11074300], Fundamental research funds for
the central university [under Grant No. 2010YS03], Technology
Support Program of LangFang [under Grant Nos. 2010011029/30/31], and
Science Foundation of NCIAE [under Grant No. 2008-ky-13].

\end{document}